\documentclass[structabstract]{aa}

\usepackage{graphicx}
\usepackage{txfonts}
\usepackage{natbib}
\usepackage[dvips]{color}
\usepackage{subfig}
\usepackage{times}
\usepackage[colorlinks,linkcolor=blue,anchorcolor=blue,citecolor=blue]{hyperref}
\usepackage[hyperref]{backref}
\usepackage{rotating}
\usepackage{longtable}
\usepackage{lscape}

\begin{document}
\title{EXSdetect: an end-to-end software for extended source detection in X-ray images: application to {\sl Swift}-XRT data}
\author{Teng Liu\inst{1}\inst{,2}, Paolo Tozzi\inst{2}\inst{,3}, Elena Tundo\inst{2}, A. Moretti\inst{4}, Jun-Xian Wang\inst{1}, Piero Rosati\inst{5}, Fabrizia Guglielmetti\inst{6} }

\offprints{Teng Liu, \email{liuteng@ustc.edu.cn}}

\institute{ $^1$CAS Key Laboratory for Research in Galaxies and Cosmology, Department of Astronomy, University of Science and Technology of China, Hefei, Anhui 230026, China\\ $^2$INAF, Osservatorio Astronomico di Trieste, via G.B. Tiepolo 11, I-34131, Trieste, Italy\\ $^3$INFN-National Institute for Nuclear Physics, via Valerio 2, 34127 Trieste, Italy \\ $^4$INAF, Osservatorio Astronomico di Brera, Via Brera 28, 20121, Milano, Italy \\ $^5$European Southern Observatory, Karl Schwarzschild Strasse 2, D-85748 Garching bei Muenchen, Germany\\ $^6$Max-Planck-Institut f¨ur extraterrestrische Physik, Giessenbachstrasse, D-85748 Garching bei Muenchen, Germany}



\abstract
{}
{
We present a stand-alone software (named EXSdetect)  for the detection of extended sources in X-ray images. Our goal is to provide a flexible tool capable of detecting extended sources down to the lowest flux levels attainable within instrumental limitations, while maintaining robust photometry, high completeness, and low contamination, regardless of source morphology. EXSdetect was developed mainly to exploit the ever-increasing wealth of archival X-ray data, but is also ideally suited to explore the scientific capabilities of future X-ray facilities, with a strong focus on investigations of distant groups and clusters of galaxies.
}
{
EXSdetect combines a fast Voronoi tessellation code with a  friends-of-friends algorithm and an automated deblending procedure. The values of key parameters are matched to fundamental telescope properties  such as angular resolution and instrumental background. In addition, the  software is designed to permit extensive tests of its performance via simulations of a wide range of observational scenarios.
}
{
We  applied EXSdetect to simulated data fields modeled to realistically represent the {{\sl Swift}} X-ray Cluster Survey (SXCS ), which is based on archival data obtained by the X-ray telescope  onboard the {{\sl Swift}} satellite. We  achieve more than $90\%$ completeness for extended sources comprising at least $80$ photons in the $0.5$--$2$ keV band, a limit that corresponds to $10^{-14}$ erg cm$^{-2}$ s$^{-1}$ for the deepest SXCS fields. This detection limit is comparable to the one attained by the most sensitive cluster surveys conducted with much larger X-ray telescopes.  While evaluating the performance of EXSdetect, we also  explored the impact of improved angular resolution and discuss the ideal properties of the next generation of X-ray survey missions.
}
{}


\authorrunning{Liu et al.}  

\titlerunning{Extended X-ray Source Detection} \maketitle

\section{Introduction\label{introduction}}

In the  past decade, X-ray astronomy reached an unprecedented level in imaging and spectroscopic performances, mostly thanks to the {\sl Chandra} and {\sl XMM-Newton} telescopes. 
 Their high imaging quality and large effective area allowed us to discover an increasing complexity in the morphology of X-ray extended sources, which reflects the rich physics involved.  
Extended X-ray emission may be associated with a variety of Galactic objects:  supernova remnants,  star-forming regions, or
planetary nebulae.
In extragalactic fields, it is mostly associated with the hot intra cluster medium (ICM) that permeates the potential well of  galaxy clusters and groups.
Other extragalactic, extended X-ray sources are given by the inverse Compton scattering from relativistic jets in radio galaxies, from  supernova remnants, hot gas and massive X-ray binaries in star-forming galaxies, and low-mass X-ray binaries in normal galaxies.
In this paper, we mainly focus on galaxy clusters and groups.

 Thanks to the development of tools for analyzing the morphology of extended sources and the capability of simultaneous imaging and spectroscopy with CCD, a wealth of new physical phenomena has been opened to direct investigation in the field of galaxy groups and clusters.
The most noticeable discoveries include the interaction between the ICM and the relativistic jets of the central radio galaxy, which is directly observed as bubbles inflated by the jets; the presence of cold fronts; and the physics of cool-cores and their metal distribution.  However, less attention has been paid to  the detection of very faint extended sources, despite the increasing interests in distant groups and clusters of galaxies.
 Based on the serendipitous discoveries of high-z massive clusters of  galaxies through X-ray observation of {\sl XMM-Newton} and {\sl Chandra} or through the Sunyaev-Zeldovich effect \citep[see][]{jee09, foley11,santos11}, we expect a significant population of massive clusters at very high redshift ($z\sim 1.5$), whose presence was considered unlikely just a few years ago. 
The detection of a  sizable number of high-z massive clusters may pave the way to a substantial revision of the standard $\Lambda$CDM cosmological model \citep{hc12,hoyle12,waiz12}.  
 Moreover, a largely unexplored population of small or medium-z groups awaits systematical study.
 Overall, a thorough investigation of the low-flux, extended X-ray sources is expected to provide significant progress in this field.

 Robust detection and characterization of faint extended X-ray sources is very difficult. 
In the photon-starving regime of X-ray images \citep{hobart2005}, Poissonian fluctuations in the sparse background can produce spurious source detections at low fluxes, which makes the firm detection of faint sources very difficult, irrespective of their extent.   In addition, detecting an extended source is much more difficult than detecting an unresolved source with the same number of net photons.  First, extended emission is spread over more pixels, resulting in a much lower contrast to the background, and can be more easily swamped by the sparse background.
Second, faint extended sources can be confused with faint unresolved sources.

In addition, the mere detection of a source is not sufficient, since a robust characterization of its extension is also required on the basis of X-ray data.  In principle, the only requirement to characterize a source as extended is that its measured size must be larger than the instrument point spread function (PSF). 
However, given the large number density of active galactic nuclei (AGNs) and distant star-forming galaxies, unresolved sources largely outnumber extended ones, especially at low fluxes.  The risk of finding faint extended emission encompassing several unrelated unresolved sources is high.  Moreover, in the classical Wolter-type design of  X-ray mirrors, the PSF varies significantly across the field of view (FOV), introducing more uncertainties.  Finally, extended X-ray sources are complex in morphology, because they come in a wide variety of shapes and surface brightness distributions, with scales ranging from a few arcsec to a few arcmin.  Therefore, a considerable effort should be made to search and characterize extended sources in X-ray images.


In this paper, we aim at providing a flexible and efficient detection algorithm to identify extended X-ray sources down to a low flux level to exploit the large amount of data in the archive of current X-ray missions ({\sl Chandra}, {\sl XMM-Newton}, {\sl Swift}, {\sl Suzaku}), and to explore the scientific cases of future X-ray facilities.  
We  mostly focus  on the detection of diffuse extragalactic sources, namely groups and clusters of galaxies. 

  The identification of extended sources consists of two parts: source detection and source characterization. 
For the first part, we use  the same algorithm -- a combination of Voronoi tessellation (VT) and friends-of-friends (FOF) -- as used in the classical software {\tt vtpdetect} based on the work by \citet{ebeling93}.
 VT is a useful tool to deal with typical X-ray images, which are largely dominated by empty pixels.
A recent application has been presented by \citet{diehl06},  who used weighted VT in adaptive binning of X-ray images.  This software is not designed for source detection, however. 

Although it has been  widely used for detecting galaxy clusters in optical images (for instance in \citet{ramella01}, \citet{kim02}, \citet{panko05}, \citet{vanbreukelen09}, and \citet{barkhouse06}) \footnote{These works identify overdensities in the field of galaxies using a third information in addition to the 2D galaxy coordinates, such as redshift, magnitude, or color.}, the method of VT+FOF is seldom used in X-ray cluster surveys.   Presently, the only X-ray source detection software based on VT available to the community is {\tt vtpdetect} as a part of the {\tt ciao}\footnote{http://cxc.harvard.edu/ciao4.4/} software developed for the {\sl Chandra} mission.  However, so far it has been applied mainly to ROSAT data \citep{scharf97,ebeling98,ebeling00}, and to our knowledge, only once to {\sl Chandra} data \citep{boschin02}.  
The detection algorithms  commonly used in X-ray surveys are based on wavelet transform methods (\citealp[the ROSAT Deep Cluster survey]{rosati98};\citealp[the ROSAT 160d survey]{vikhlinin98_160d};\citealp[the ROSAT 400d survey]{burenin07_400d}; \citealp[the {\sl XMM-Newton} Distant Cluster Project]{fassbender11_XDCP}; \citealp[the XMM Cluster Survey]{lloyd-davies11_XCS}; \citealp[The XMM Large-Scale Structure survey]{pacaud06}; \citealp[the {\sl Chandra} Multi-wavelength Project]{barkhouse06}).

Very few alternative lines of research are being developed for exploiting the entire information contained in X-ray images.  So far, the only method fully employing the information of the photon counts per pixel and of the counts in neighboring pixels is the background/source separation technique \citep{guglielmetti09}.  
 This method uses the Bayesian probability theory combined with a two-component mixture model.  
This way, background and source intensity can be  estimated jointly.  First results based on this technique and applied to deep {\sl Chandra} fields are currently under scrutiny (Guglielmetti et al. in preparation).
 Given this framework, it is well worth putting a significant effort into developing diversified detection algorithms for extended sources.



 In this paper, we further develop the VT+FOF algorithm, making a great effort against its major shortcomings.
We provide the flexible and efficient {\sl Extended X-ray Source detection} software ({\sl EXSdetect}), which can be easily applied to a wide range of X-ray images, and is optimized to detect extended sources. 
The software, written in {\sl Python} and made available to the community on a public website, can be  used for a double purpose: exploiting the rich (and increasing) archive of current X-ray missions; and investigating scientific cases of the next-generation X-ray facilities, in particular those that will perform large-area surveys. 
Here we do not consider supplementary information, although we may use the CCD spectral information to distinguish thermal (X-ray soft) from nonthermal (X-ray hard)  emission.  
 This option is not practical, because mostly the signal-to-noise ratio (S/N) is very low in the hard band (above 2 keV) where the difference between thermal and nonthermal spectra is highest.  
We also ignore the possibility of cross-correlating X-ray images with images in other wavebands like optical and IR.  This process is very effective in finding distant cluster candidates, but it would severely undermine the possibility of clearly defining the surveyed volume, an essential requirement to derive the physical density of the cluster population as a function of the cosmic epoch.   Our goal is to obtain samples selected entirely by their X-ray properties and characterized by clearly defined completeness criteria.  This requirement is mostly relevant for cosmological tests, where we need a robust estimate of the search volume of the survey to measure the comoving number density of sources.

The paper is organized as follows.  In \S\ref{algorithm} we review the {\tt vtpdetect} algorithm and describe the improvements and the additions of {\sl EXSdetect}.  In \S\ref{simu} we discuss the practical case of the ongoing {\sl Swift} X-ray Cluster Survey \citep[SXCS,][]{tundo12}, presenting extensive simulations aimed at testing the performance and evaluating the completeness of the survey.  In \S 4 we briefly discuss the effect of improving the angular resolution.  Finally, in \S\ref{conclusions} we summarize our findings.

\section{Algorithm\label{algorithm}}
\subsection{ VT+FOF as implemented in {\tt vtpdetect}\label{VTP}}
There are several ways to find photon density enhancements, and therefore identify source candidates in an X-ray image.  
The simplest way is to search across the image inside sliding boxes with different sizes and shapes.   
A more effective option is to search for overdensities on different scales in Fourier space,  making use of wavelet transform.
Another possibility is to calculate the local flux density across the image in Voronoi cells and locate excess in the flux density distribution.  
This method has been introduced by \citet{ebeling93} and is implemented in the {\tt ciao} task {\tt vtpdetect}.  
 This algorithm is a combination of the classical VT and FOF.  
The classical VT is a nonparametric method that uniquely partitions the infinite plane into convex polygon cells according to a list of starting locations (named Voronoi sites) in the plane.  
Each cell contains only one Voronoi site and includes all points that are closer to this site than to any other. The positions of all occupied pixels are assumed as Voronoi sites, and the local photon density value associated to each occupied pixel is computed as the number of photons in the occupied pixel divided by the Voronoi cell area.
 In this way, the raw image is converted into the local photon density map, in which high-density regions show up by a direct comparison between the photon density map and a background map.
Then, the FOF method is used to identify source candidates.

The VT+FOF method has two immediate advantages.   First,  it is applied to the X-ray photon event list or to the original unbinned image, and it preserves the full angular resolution at each step, at variance with other methods, where image-binning is often adopted to avoid oversampling of the PSF and improve the low count statistics.  
Second, the VT+FOF algorithm does not require assumptions about the shape and size of the structure one is looking for. 
 In the simplest approach, a measure of the size (area) of the source candidate and the corresponding aperture photometry is directly obtained  by defining the source edges as the {\sl loci} where the photon density value is equal to the background density, without using a predefined aperture region.

Despite these properties, {\tt vtpdetect} is not widely used compared to wavelet-based algorithms.
The construction of VT has been considered to be very time-consuming, and {\tt vtpdetect} was suggested to be applied only to small fields or low-event-density regions \citep{ebeling93}.  Another major reason for this infrequent use is that the FOF algorithm has a severe source-blending problem, i.e., the merging of two or more neighboring sources into one.  This is a well-known problem, which is particularly severe in deep fields where both the background and the source number density are high.  Therefore, using {\tt vtpdetect} requires significant additional work to obtain a well-characterized list of extended sources.

 In this work we present a new end-to-end software by implementing the same VT+FOF method as used in \citet{ebeling93} and solving these drawbacks.  
Most importantly, we add a self-contained deblending procedure and provide a well-characterized extended source list as the output.
We preserve all advantages of {\tt vtpdetect}, without introducing any image binning or smoothing, or assuming any specific source shape or size.
The details of our algorithm are given in the following sections.


\subsection{Voronoi diagram construction\label{VT}}
The pattern of the Voronoi cells is also called a Voronoi diagram,  which is composed of the edges of the cells (Voronoi edges).   A simple algorithm for constructing a Voronoi diagram at the highest computational speed is the {\sl sweep-line} algorithm \citep{Fortune1986}
\footnote{The computing time scales as $O(n\log n)$, where $n$ is the number of occupied pixels. Using the sweep-line algorithm, the time spent on Voronoi construction is negligible compared to the subsequent steps.}. 
 We implemented this algorithm and further developed it for better and faster performance.  
As a first step, we confined the Voronoi diagram, originally defined in the infinite plane, to the box domain of the image.
Second, in addition to the classical Voronoi diagram, we built a discrete Voronoi diagram by including all pixels enclosed by a Voronoi cell into a discrete Voronoi cell.  If one pixel is equally distant from two or more Voronoi sites (i.e., the pixel center falls on an Voronoi edge), it is assigned randomly among the two or more cells.
Finally, as a third step, we connected the pair of Voronoi sites on both sides of each Voronoi edge, building the so-called Delaunay diagram, which will constitute the structure on which FOF and our deblending procedure are run.

\begin{figure*}[htbp]
\includegraphics[width=\textwidth]{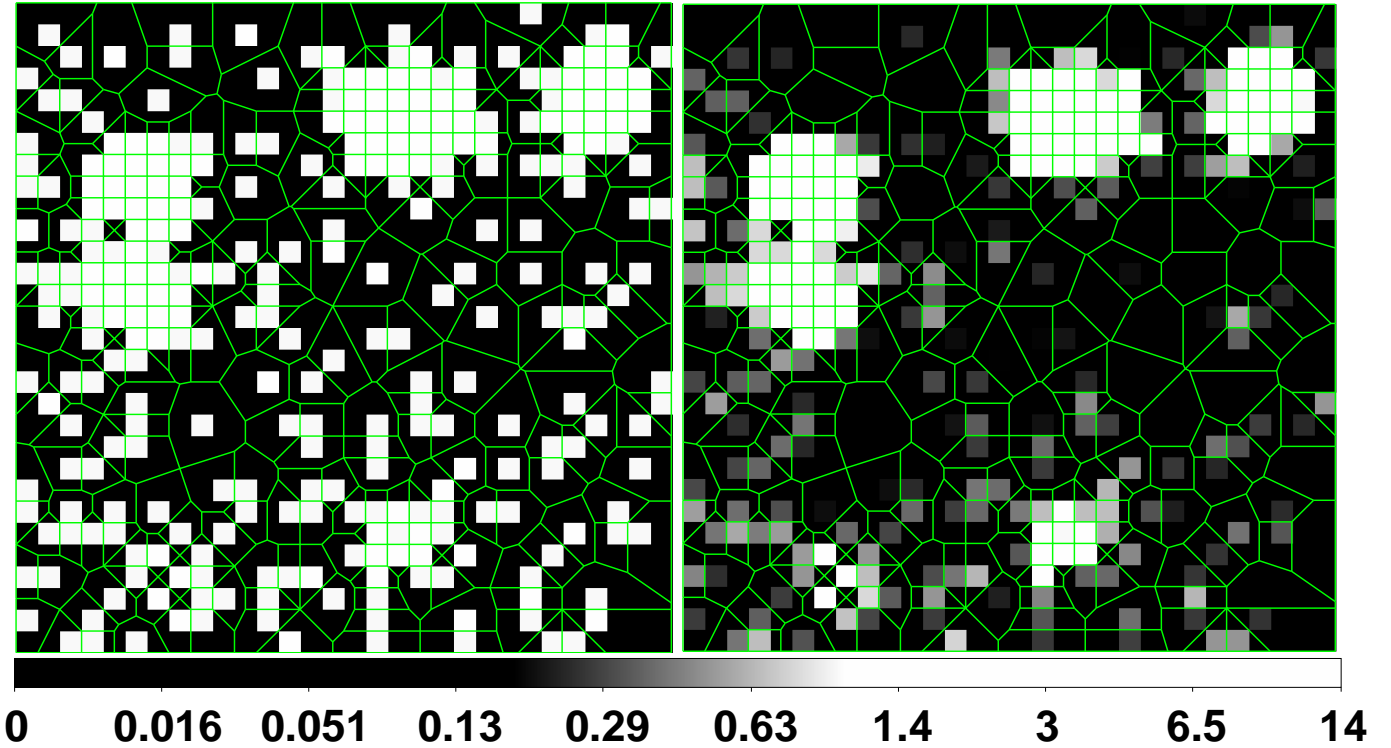}
\caption{Left panel:  original unbinned X-ray image with the
  Voronoi tessellation (green lines) obtained with the classical
  sweep-line algorithm.  The value of each filled pixel is simply the
  number of photons.  Right panel: photon density map obtained  by
  dividing the original image by the area map. The intensity of each
  filled pixel is now proportional to the photon density.}
\label{opdCorr}
\end{figure*}

Based on the Voronoi diagram, we created an area map that contains the area  of Voronoi cells associated to each occupied pixel (Voronoi site).  We calculated the accurate area of the cell polygons rather than counting the pixels in the discrete Voronoi cells.  Finally, we computed the photon density map by dividing the original image by the area map.  An example of a Voronoi tessellation and its associated photon density map is shown in Figure \ref{opdCorr}. 
 To identify source candidates, the photon density map must be compared with a background map, which is generated as described in the following section.

\subsection{Background map generation\label{bkgmap}}


 A photon in an X-ray image can be associated to a source or the background.
The background is defined as the sum of all contributions not associated to astronomical sources, or associated to some astronomical component that cannot be resolved (like the Galactic diffuse emission). 
We estimated the average background flux adopting the same method as used in \citet[][see their section D]{ebeling93}.  
We assumed that the background photons in the exposure-corrected image, which is obtained by dividing the photon density image by the exposure map, are distributed according to the Poisson statistics, and that the fraction of filled pixels with the lowest photon density values are dominated by background. Fitting a simple analytical model to the distribution of the low photon-density pixels, an average background intensity was estimated. The final background map of the image in units of $photon/pixel$ is obtained by multiplying the mean background flux by the exposure map.  In this way the variations across the field of view due to vignetting, edges of the chips, or missing columns and pixels were taken into account.  

We adopted a constant background for each exposure-corrected XRT image.
This choice is reasonable for XRT images as we tested with simulations.  However, it may not apply to images with strong background variations.   Another caveat is that this method works properly when the number of occupied pixels is dominated by  background photons.  Thus it cannot be applied to fields dominated by source emission (for example, very shallow fields with very bright sources, or with extended sources that cover the entire FOV).  
For such fields a manual treatment of the background is preferable.  We plan to introduce a more  sophisticated background treatment in a future version of the code. 



\subsection{FOF detection and deblending\label{fof}}

In the FOF algorithm two occupied pixels whose distance is smaller than a chosen linking length are defined as {\sl friends}.  Starting from any occupied pixel, all its {\sl friends} are associated with it, and then {\sl friends} of its {\sl friends}, until no more new {\sl friends} can be found.   All connected pixels (or the corresponding Voronoi cells) constitute a source candidate. The FOF algorithm is iteratively applied to all involved pixels, until all of them are assigned to some source candidates.  We ran FOF on the occupied pixels whose photon density was above the background level.  This choice is very generous in the sense that several spurious detections may survive above the background level. However, we preferred to start with a large list of candidates  and to refine it later, a procedure often adopted \citep[see ][]{broos}.

The direct application of the FOF algorithm results in a large number of blended sources, as mentioned also in \citet{ebeling93}.  However, given the relatively low number density of sources in the X-ray sky
\footnote{This clearly depends on the depth of the image.  At the flux levels currently explored in the deepest survey \citep[$\sim 10^{-17}$ erg cm$^{-2}$ s$^{-1}$ in the soft band in the CDFS, see][]{xue11}, this value is about $2\times 10^4$ deg$^{-2}$ \citep{lehmer2012}, and only thanks to the high angular resolution of {\sl Chandra} this value is still far from the confusion limit.  Nevertheless, most of the archival X-ray data available at present are, at best, one or two orders of magnitudes less sensitive than the CDFS, which will remain for a long time the deepest X-ray image of the sky.}, most of the blended sources overlap only in their fainter outer regions, while their bright cores, corresponding to the peak of the PSF, remain isolated.  
In principle, the blended sources can be deblended by raising the threshold well above the background level.  

The threshold necessary to separate two blended sources varies according to the background, the distance between the sources, and the size of the sources themselves, which, for unresolved sources, is determined only by the PSF profile.   
This step has been performed by \citet{horner08}  on the {\tt vtpdetect} direct output.  
They ran {\tt vtpdetect} five times on each X-ray image using different thresholds  of surface brightness, and then selected the best threshold for each field by visually inspecting the source photon distribution for each threshold value.  
This procedure  is reliable but time-consuming, especially for a large survey that spans a wide range of exposure times and therefore surface brightness thresholds.  
As previously anticipated, a major improvement of our new software consists in including the auto-deblending procedure in the algorithm.  

\begin{figure}[htbp]
\includegraphics[width=\linewidth]{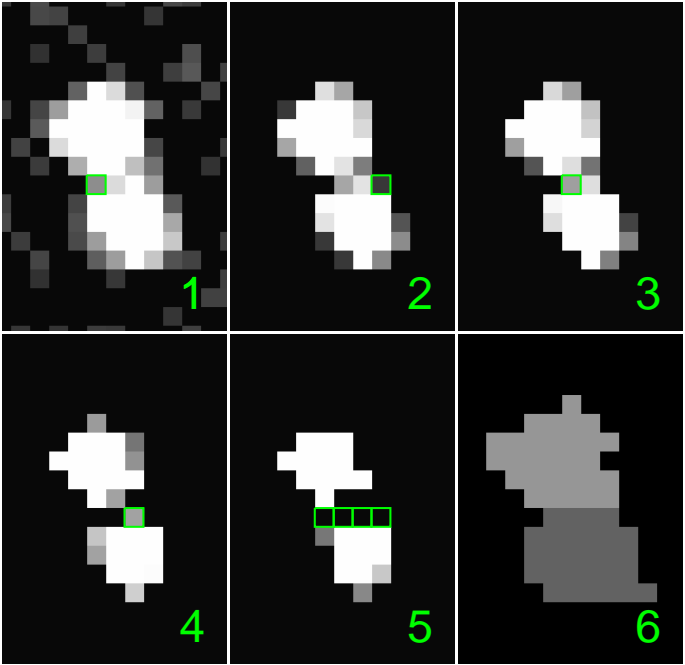}
\caption{Example of the deblending procedure applied to two unresolved
  sources spaced by $\sim 6$ pixels.  In panels 1--5 we show the pixels
  whose photon density is above a threshold that increases at each step.
  {\sl Bridge pixels} (see text for the definition) are marked with
  green boxes and are identified backwards starting from panel 4 to 1.
  Panel 6 shows the two sources identified separately, with the {\sl
    bridge pixels} re-assigned according to a local criterion.}
\label{separate}
\end{figure}

 As a first step, to flag blended sources, we created a number of source maps\footnote{The number of the maps can
  vary; a number around 10 guarantees robust results.} by running FOF with thresholds corresponding to different photon densities, starting from the lowest (corresponding to the background level, map with index $j=1$) to the highest value present in the photon density image, distributed on a logarithmic scale. 
Clearly, the source regions defined in the map $j=1 $ will be larger than the regions in the higher order maps. 
An example of this process is shown in Figure \ref{separate}, where two blended sources are identified by FOF as one in panels 1 to 4, and finally as two separate sources in panel 5.  
A source is tagged as blended whenever its source region includes more than one source in one of the higher order maps.

 At this point we had a list of sources virtually free from blending.  Their positions were assigned by searching for the pixel whose $3\times 3$ square island has the largest number of photons.  Then we separated the blended sources among the most reasonable non-overlapping source regions, without assuming a source brightness model.  
This is obtained as follows.  For each pair of blended sources, we can find a threshold above which they are detected without overlapping pixels.  Sorting all occupied pixels by  their photon densities, we identify the pixel with the lowest value whose inclusion causes the bridging of the two sources.  This pixel is flagged as a {\sl bridge pixel} and is removed from the original image.  After a {\sl bridge pixel} is found and removed, we again run FOF on all occupied pixels above local background.  Unless the sources are detected separately by FOF, we repeat the above procedure to identify a new {\sl bridge pixel}.  The final set of {\sl bridge pixels} is the minimum set of the faintest filled pixels we need to remove to separate the two sources.  This process can be seen in Figure \ref{separate} going from panel 5 back to panel 1.  As soon as the set of {\sl bridge pixels} are found and removed, the regions of the two sources  are built directly with FOF.  At the last step, each one of the bridge pixels are re-assigned to one of the two sources, with a simple criterion based on the  neighboring filled pixels: if the majority of them belong to a given source, the {\sl bridge pixel} is assigned to this source.  In the very few cases where this condition does not apply, the bridge pixel is assigned randomly.  The final result is shown in panel 6 of Figure \ref{separate}.

 There are some cases that require  special treatments.  The first one concerns blending of extended sources. In principle, blended extended sources cannot be separated without making assumptions on their surface brightness distributions.  However, for most of the extended sources we are interested in, such as regular clusters with only one bright cool core, it is reasonable to assume that the surface brightness decreases monotonically from the core to the edge.  In such cases, blended extended sources can be separated with the same method as described  above.  
This step is left as an option.  The second case is the bridging between extended sources and background overdensity regions, which results in branch-like shapes at the edge of extended sources.  We managed to clean such cases by identifying the {\sl cut-vertex}.   A {\sl cut-vertex} is similar to a {\sl bridge pixel}, but it is defined as the single pixel whose removal allows one to separate the blended sources into two or more components in the image.
Clearly, the presence of a {\sl cut-vertex} indicates a weak connection between two regions.  Cutting off these weak connections allowed us to efficiently reduce the bridging of real extended sources with background fluctuations.

\subsection{Reliability filter and source classification\label{classification}}

\begin{figure}[htbp]
\includegraphics[width=\linewidth]{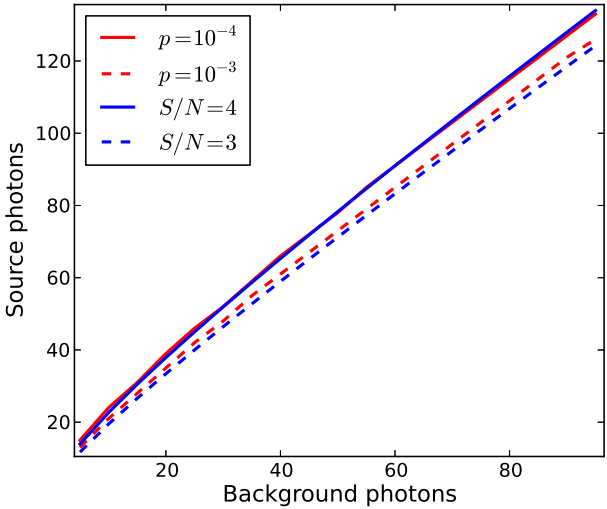}
\caption{Curves in the net photons-background photons plane
  corresponding to a probability $p$ of $10^{-3}$ or $10^{-4}$ of being a
  random fluctuation (red lines), compared with those corresponding to
  $S/N$ of $3$ or $4$ (blue lines).  Curves for $S/N=4$, $p=10^{-4}$
  and $S/N=3$, $p=10^{-3}$ are shown with solid and dashed lines,
  respectively.}
\label{poisson}
\end{figure}

 We applied a reliability criterion by comparing the total photons $C_{total}$ inside the source region to the number of background photons $Bkg$ expected in the same area, as obtained from the background map.  
We computed the S/N as $C_{net}/\sqrt{Bkg}$, where $C_{net}=C_{total}-Bkg$.  
This is not the traditional definition of the S/N, which is net counts over the square root of the total counts 
(source plus background) for Poisson statistics.  
However, this definition\footnote{See also http://heasarc.gsfc.nasa.gov/W3Browse/all/cxoxassist.html} is a good estimator of the probability of a source to be inconsistent with a Poissonian background fluctuation.  
In Figure \ref{poisson} we show the curves in the $C_{net}-Bkg$ plane that correspond to $S/N = 3,\, 4$ and to a Poissonian probability of $10^{-3}$ and $10^{-4}$ of to be a random fluctuations.  
 A filter for a given reliability threshold can be applied by setting the S/N threshold to the required confidence level. 
In the remainder of the paper we set this threshold to $S/N > 4$, corresponding to a Poissonian probability of $10^{-4}$ of the signal inside the source region being random fluctuation.
We verified {\sl a posteriori} that with this condition our algorithm is able to detect extended sources down to a flux level much lower than that required to characterize them as extended.
Therefore, this step does not significantly affect the final flux limit for extended source detection.
 The filter was applied not only to the final source list, but also {\sl on the run} to the temporary source list to avoid applying the extension criterion to unreliable sources and thus to reduce the computation time.

\begin{figure*}[htbp]
\raisebox{0.7cm}[0cm][0cm]{\includegraphics[width=0.4\textwidth]{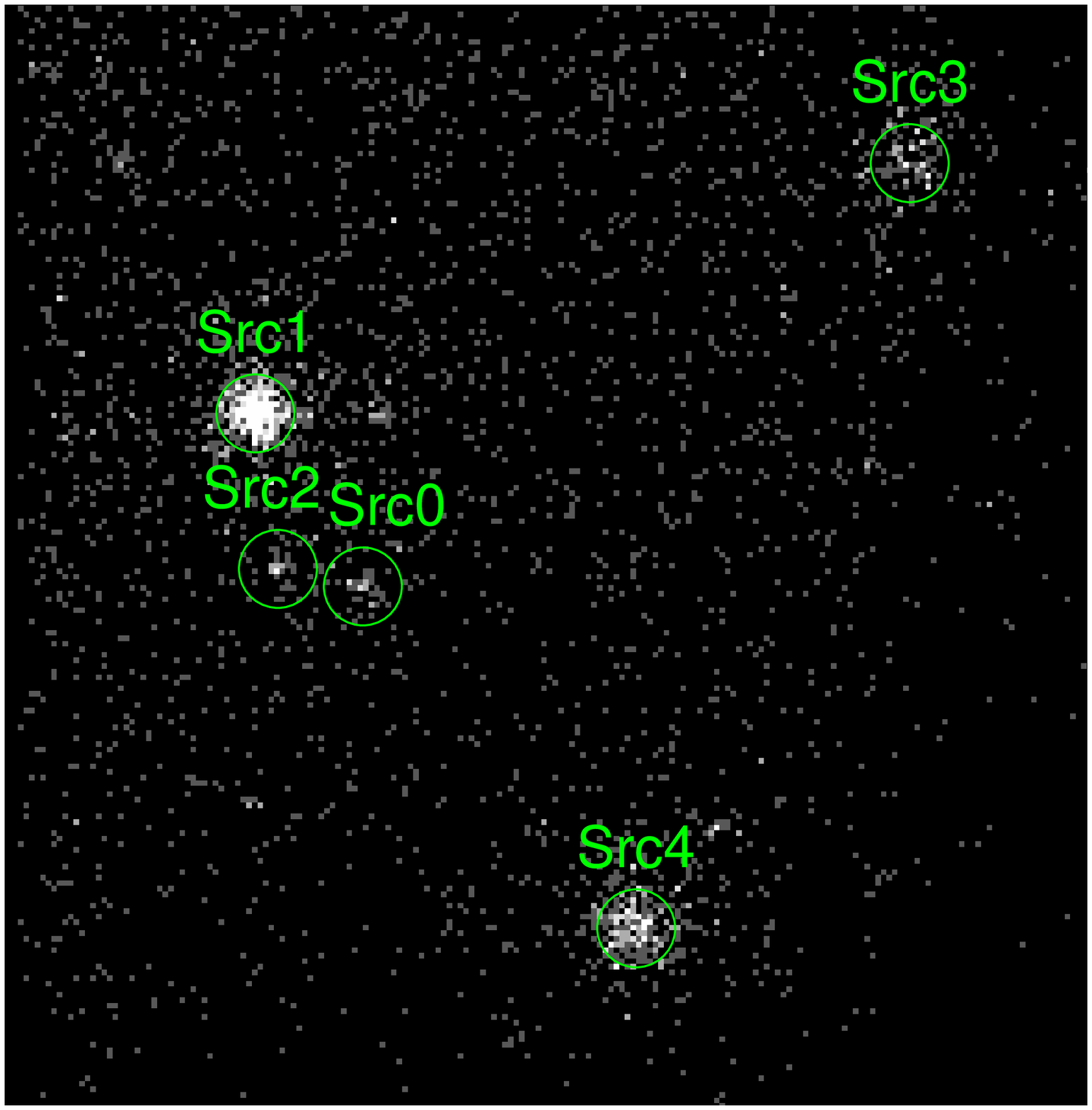}}
\includegraphics[width=0.65\textwidth]{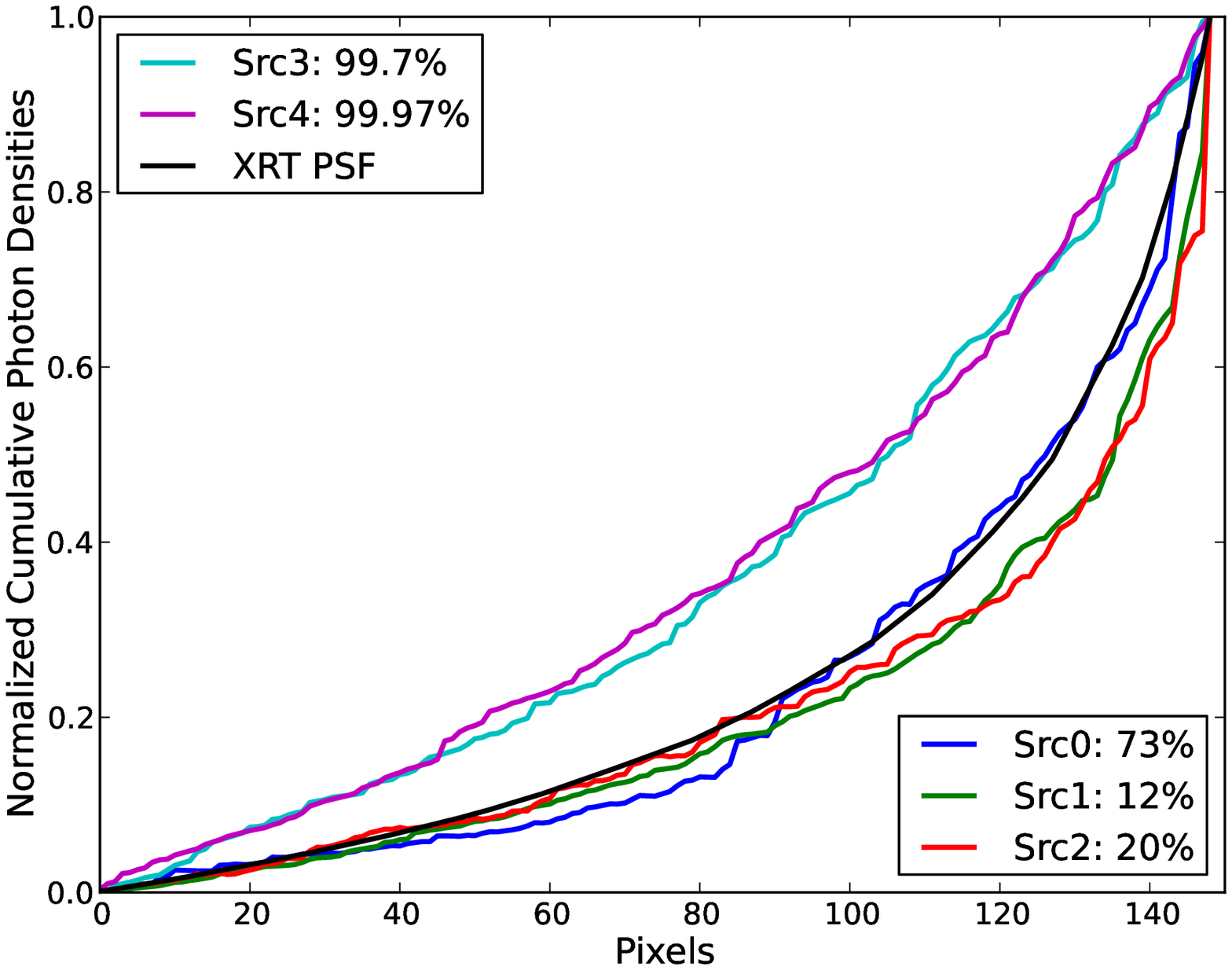}
\caption{Left panel: a simulated XRT image with five labeled sources,
  three AGNs (sources  0, 1, and 2) and two clusters (sources 3
  and 4).  Right panel: the cumulative net count profile for the
  sources shown in the upper panel. The KS-test results are given for
  each source in the inset frame.  The two clusters are clearly
  identified as extended (with a profile above the PSF and a high
  KS-test probability) and the three AGNs as unresolved sources (with
  a profile consistent with that of the PSF and a low KS-test
  probability). }
\label{kstest}
\end{figure*}

To classify all sources according to their extent, we drew a circle centered on the position of each source after masking all surrounding detected sources.  The radius was chosen to enclose $60\%\sim70\%$ energy  (slightly larger than $HEW/2$ of the PSF).  
For the sources that survived the reliability filter, we produced a normalized cumulative profile in the core circle by sorting each pixel according to the distance from the source position, and summing all photons at shorter distance
\footnote{For an asymmetric PSF the distance of each pixel to the source position can be changed to the according PSF value of each pixel.}.  A reference profile for the instrument PSF was calculated in the same way.  
The two profiles were then compared with a Kolmogorov-Smirnov (KS) test.  
Under the null hypothesis that they are drawn from the same distribution, this procedure gives us the probability that the source image is not consistent with being unresolved in the core region.
An example is shown in Figure \ref{kstest}.  Based on this procedure, we can classify the source as unresolved or extended  by setting a threshold on the null-hypothesis probability of the KS test.  
This threshold must be chosen {\sl a posteriori} after extensive simulations to keep both the contamination and the completeness of the sample  under control.  
This is shown for {\sl Swift}-XRT in \S \ref{simu}.

The KS null-hypothesis probability also depends on the net detected photons and the background level.  As we show in Section 3, one simple but effective way is to set the threshold that divides the unresolved from the extended source in the probability-S/N plane.  The optimal threshold is determined through extensive simulations by computing the completeness (fraction of recovered extended sources) and the contamination (number of sources spuriously classified as extended) and choosing the best compromise.

One may argue that a small region (the core circle) is not the best choice for classification, because extended sources are maximally different from unresolved sources at large radii.  However,  because of the low S/N in the outer regions, it turns out that it is more efficient to focus on the core region where the S/N is maximized.  However, this choice
may cause the loss of extended sources whenever a bright unresolved source in the center is present.\footnote{This effect seems to be negligible in real X-ray images, as shown, for example, by the X-ray follow-up of optically and IR selected clusters,  which indicate that the occurrence of strong X-ray unresolved sources associated with the extended emission from groups and clusters of galaxies is  very rare \citep[see ][]{hicks08,big08}}.  This may occur because of
a strong (typically unrelated) unresolved source is embedded in the extended emission, or because of the presence of a bright cool core whose size is below the angular resolution.  To overcome this problem, we applied a supplementary step to recover most of these sources.  We considered all sources that are classified as unresolved but with KS null-hypothesis probability $> 50\%$.  For each of these sources, we normalized the local PSF to the total photon inside the core region and calculated the radius where it reaches $1/2$ the {\sl local} background level.  
Since the edges resulting from the FOF algorithm are set by the {\sl overall} background level, the circles defined
by this criterion should not be entirely encompassed by the FOF source regions.
If this happens, though, it may imply the presence of extended emission embedded in a larger background region (for whatever reason) and the source is tentatively classified with a flag as extended.   We remark, however, that this procedure may introduce several false candidates if the PSF shape is not completely under control.  Eventually, these cases may be  visually inspected for final inclusion in the extended source list.  Finally, as a very last step, extended sources found within a larger, surrounding extended source are merged into it and considered part of the largest one.  

\subsection{Source regions and aperture photometry\label{photometry}}

 The definition of a FOF detected source region depends  on the linking length, which is the only parameter of FOF.  The linking length is defined as the median distance between occupied pixels in a pure background image, and therefore it depends on the background intensity  only.  An estimate of the linking length in units of pixel size is given by $\sqrt{Bkg^{-1}}$, where $Bkg$ is in units of $photon/pixel$.  Clearly the linking length depends on the position on the image, therefore we defined a linking length map according to the background map.  
 Using this linking length, the borders of each source are automatically defined as the {\sl loci} where the photon density is equal to the background density.  We recall that the photon density still includes both the source signal and the background, since no background was subtracted from the image.  With this definition, statistically the entire signal associated to the source is included in the FOF region. In other words, this choice of the linking length corresponds to a precise definition of the source boundaries, and allows us to perform aperture photometry directly inside the FOF region. 

 For the sources classified as extended, the regions obtained by FOF were used as photometry aperture. 
However, unresolved sources that are well-described by the PSF shape were treated differently.
For each source classified as unresolved, we normalized the PSF to the total photon inside the core region and calculated the radius where it reaches $1/2$ the local background level.  This defines the extraction radius for all unresolved sources.   When two unresolved sources are closer than this radius, the overlapping pixels were assigned to the one with
the brighter expected value in that position. 
A more sophisticated treatment including a source-fitting procedure is planned to be introduced in a future version of the algorithm.

We calculated the total number of photons within each source region from the original image, and the total background photons within the same region from the background map.  Aperture photometry for each source was simply computed as the total number of photons minus the number of background photons.  After all these steps, we obtained a list of extended source candidates without any {\sl a priori} assumptions on their intrinsic size and properties.  
For each source, we provide position and area, together with  its aperture photometry, and the normalized exposure time corresponding to the emission-weighted exposure in the source region.


\section{Application to {\sl Swift}-XRT archival data\label{simu}}


One of our goals is to provide a flexible algorithm that can be used for extensive simulations adapted to specific cases.  
 This is a relevant aspect since the performance of any algorithm changes significantly as a function of instrument characteristics, such as resolution and background level.  In this section we describe set-up of simulations to obtain a characterization of a survey in terms of completeness and contamination.  
We describe the specific case of the X-ray Telescope (XRT) onboard the {\sl Swift} satellite, which has been used mainly to follow up gamma-ray bursts (GRB), and which so far provided a serendipitous survey of about 70 deg$^2$ of the X-ray sky
\citep[see][]{tundo12}.  The result of this section will be used in a forthcoming paper to characterize the final catalog (from about 400 deg$^2$) of  SXCS (Liu et al. in preparation).

\subsection{XRT image simulation} 

\begin{figure*}[htbp]
\includegraphics[width=\textwidth]{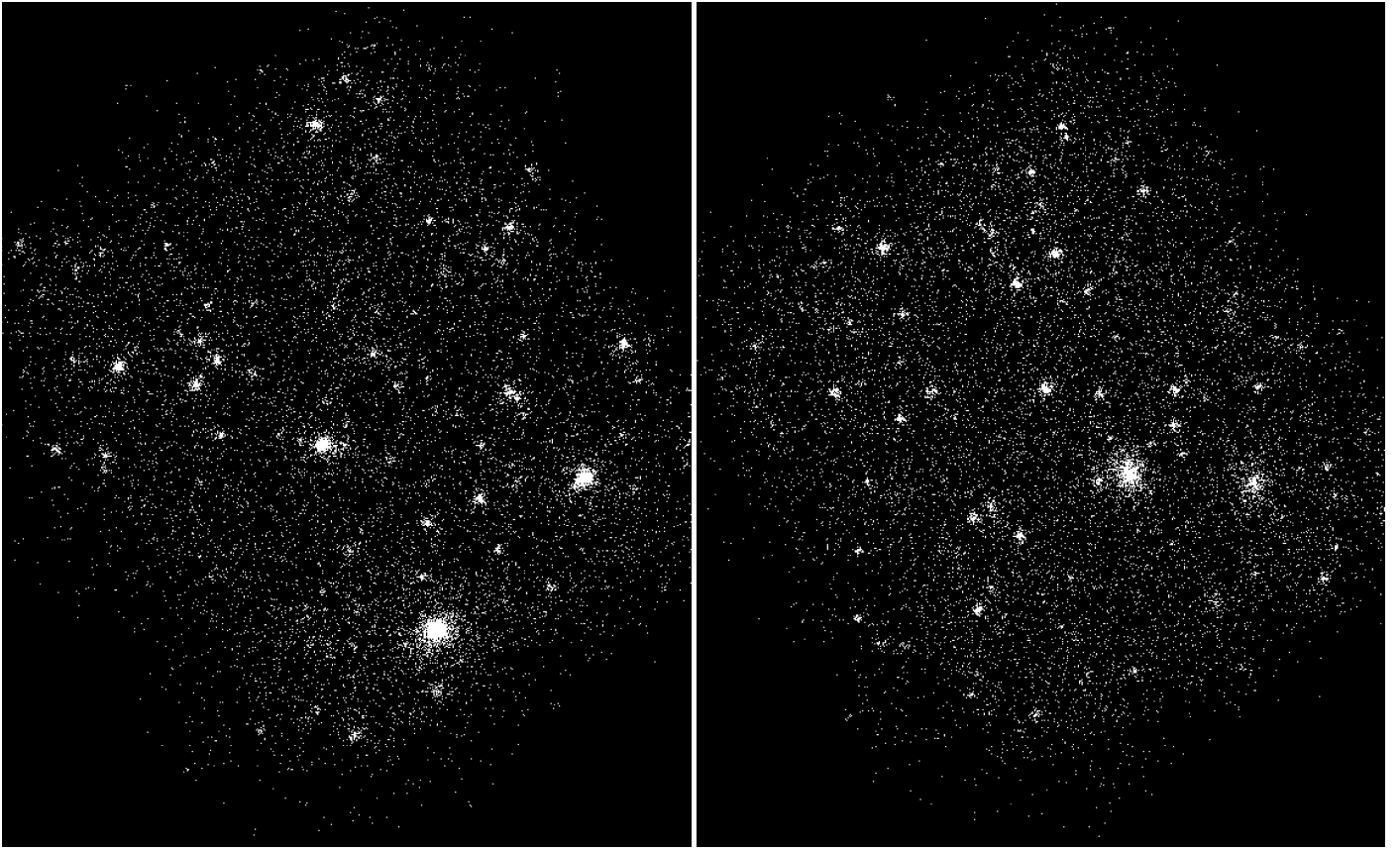}
\caption{Soft band (0.5--2.0 keV) image of the field 110059+514249
  from the XRT archive (left), and a mock image  simulated for the same field.}
\label{realmockimages}
\end{figure*}

\citet[][]{tundo12} used a first chunk of the XRT archive as of April 2010 including 336 GRB follow-up fields with Galactic latitude $|b|>20\ $deg, resulting in a catalog of 72 extended X-ray sources  (catalog I).
 Several of these sources have already been identified as groups or clusters thanks to a cross-correlation with available optically data (mostly from the Sloan Digital Sky Survey) or have been confirmed as clusters thanks to a detailed X-ray spectral analysis (Moretti et al. in preparation).

The selection method used in \citet[][]{tundo12} was based on the {\tt wavdetect} algorithm and a simple growth-curve characterization.
Completeness and contamination levels  were kept under control thanks to the relatively high detection threshold adopted in catalog I (corresponding to at least 100 net counts in the soft band).  
We intend to increase both the sensitivity and the sky coverage of  SXCS by applying the algorithm described here to the full {\sl Swift}-XRT archive (updated to November 2011), after removing all fields with Galactic latitude $|b|>20$ deg or originally targeted at groups or clusters.
 Thanks to a much larger number of fields ($\gtrsim 3000$) and a lower flux limit than that adopted in \citet[][]{tundo12}, we expect a significantly larger number of extended sources ($\gtrsim 300$).  
To push the detection limit of extended sources down to the faintest possible fluxes, we need to robustly assess the completeness and the contamination levels of the survey.

 To the aim of building simulations as close as possible to the actual survey, we generated one image for each field in the soft band using  the same exposure time, exposure map, Galactic column density, energy conversion factor (ECF), and background flux as in the real image (see Figure \ref{realmockimages} for a comparison of a real image and a mock image of the same field).   Point sources were randomly extracted from a distribution consistent with the  logN-logS measured in deep {\sl Chandra} fields \citep{2002Rosati,lehmer2012} and simulated down to fluxes about one order of magnitudes lower than the expected detection limit of XRT images.   For each input source we converted the intrinsic soft-band flux into photon rate, using the appropriate ECFs, which are computed as in \citet[][]{tundo12}, taking into account the Galactic absorption.  While computing the ECF, we assumed a spectral slope of $\Gamma = 1.4$, which is justified by the observed stacked spectrum of all unresolved sources in the flux range of interest, as shown in deep-field studies \citep[see, e.g.,][]{tozzi01}.  Given the exposure time, we computed the expected photon rate from each simulated source.  Then the image of each unresolved source was created by distributing these photons according to the PSF of XRT \citep{2007Moretti}.  After randomly distributing the sources across the field, the exposure map was used to apply vignetting correction at a given position.  

It is admittedly a limitation of our simulations that we assumed no spatial correlation between sources.  However, we argue that adding a correlation among X-ray sources would not significantly affect the final results.  In particular, our algorithm includes a treatment of unresolved sources embedded within extended ones (see the last paragraph in \S \ref{classification}), which reduces the impact of the correlation between unresolved and extended sources. 

The background level of each field was measured from the actual SXCS image with the method described in Section \ref{bkgmap}.  This background was added with a Poissonian distribution to the mock image.   
The added background already includes the contributions from sources below the
detection limit of each image.  
Since the mock unresolved sources are simulated down to fluxes well below the detection limit in XRT images,
the contribution of sub-threshold sources would be counted twice.  
To account for this effect, we consistently revised the value of the background downward.  
The background values were checked  a posteriori by comparing the background of the final simulated images with that in the corresponding real images.

The flux distribution of the input extended sources was taken from the logN-logS of groups and clusters measured in the ROSAT deep cluster survey \citep{rosati98}.  
We did not consider other extended sources such as radio jets or nearby resolved normal galaxies.  
As previously mentioned, the morphology of extended sources may affect detection at a significant level toward the faint end.
To take into account the different morphologies of groups and clusters, we used ten real images of relatively bright groups and clusters of galaxies obtained with the {\sl Chandra} satellite, covering a wide range in
ICM temperature (from 2 to 8 keV).  The source image was obtained by rescaling the original {\sl Chandra} image to the number of net photons expected for XRT, and to the source size typically corresponding to the source flux, adopting the phenomenologically observed relation between X-ray size and flux.  This is just an approximation adopted to avoid unrealistic sources with large size and low flux or vice versa, but the details or the intrinsic scatter of this relation do not affect the results of the simulations.  This technique has already been used to investigate the evolution of cool core clusters at high-z in \citet{santos10}. The expected net photons were computed as for the unresolved sources, adopting the ECF appropriate for a thermal spectrum, which, as shown in \citet[][]{tundo12}, has a weak dependence on the actual temperature and redshift of the source.  Finally the mock cluster image was convolved with the XRT PSF.

\subsection{Detection efficiency and contamination}

\begin{figure}[htbp]
\includegraphics[width=\linewidth]{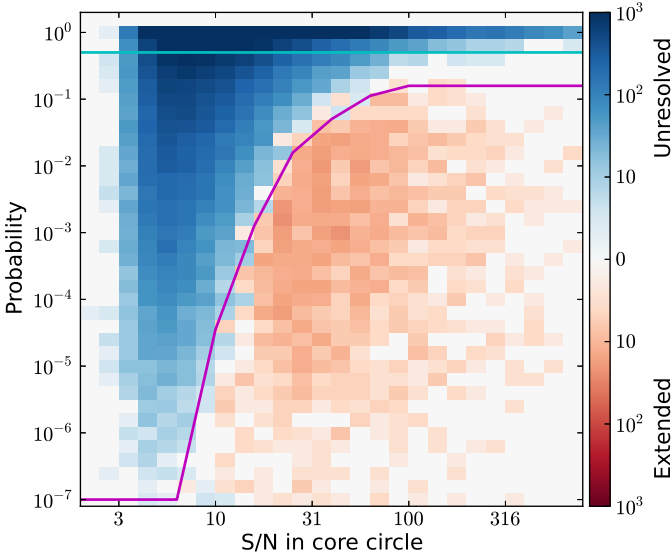}
\caption{  Selection curves in the KS-test probability--S/N plane (magenta solid line).  
The color-coded grid indicates the  absolute value of the number of unresolved minus the number of extended sources.  Regions dominated by unresolved sources are depicted in blue, while regions dominated by extended sources are shown in red.
The selection curve has been defined as the loci of  zeros on the grid.  The cyan line refers to a KS-test probability of $50\%$. }
\label{distinguish}
\end{figure}

According to the definition in \S \ref{classification}, a core radius of 5 pixel sizes (corresponding to $\sim 12$ arcsec, enclosing $\sim60\%$ of the total energy) was chosen.  In Figure \ref{distinguish} we show the threshold in the KS-test probability--S/N plane that we used to classify a source as extended.  In the same plane we show the difference between the number of extended and unresolved sources  as a color-coded grid.  This immediately shows that the threshold has been defined simply by the condition of having an equal number of spurious and extended source in a given position on the grid.  In the following we compute the fraction of recovered extended sources and the contamination level of our survey based on this simple choice.  We stress that the final properties of the survey critically depends on this selection curve.  One may want to change the selection curve to find the best compromise between completeness and contamination required for a given scientific goal: this can be done by applying different selection curves to the simulations.

\begin{figure}[htbp]
\includegraphics[width=\linewidth]{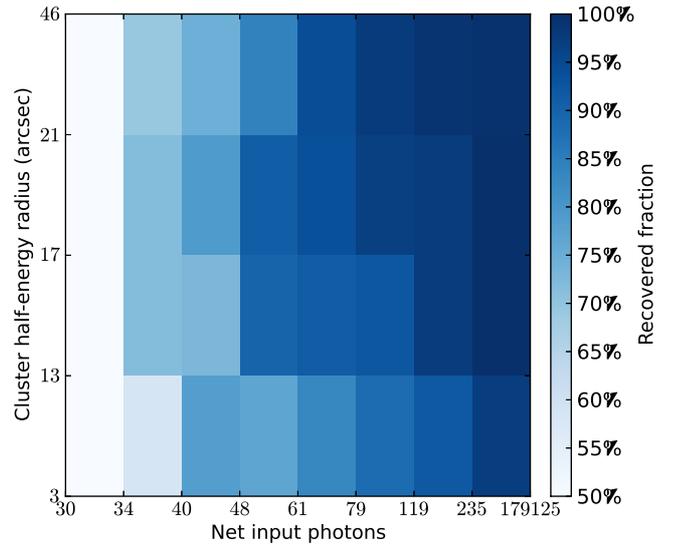}
\caption{Color-coded 2D completeness distribution in the photometry-size plane. For a given value of detected net photons, larger sources are more easily identified (i.e., detected and characterized as extended). }
\label{completeness}
\end{figure}

The completeness is defined as the fraction of recovered extended sources  in a given net photon bin.
We have a $100\%$ completeness above $\sim 200$ input net soft photons, and a gentle decrease down to $80\%$ at $\sim 60$ photons.  As shown in Figure \ref{completeness}, the completeness also depends on the source extent, with larger sources having higher identification probability.  This effect of course depends on the cluster population  we used, which may not reflect the actual cluster population at high-z (or faint fluxes) that is still largely unknown.  
In the following,  when we consider the average completeness as a function of the source photometry, we  account for the mix of surface brightness  distributions used in the simulation.
We acknowledge that a group/cluster population with different morphologies can give different completeness curves.

\begin{figure}[htbp]
\includegraphics[width=\linewidth]{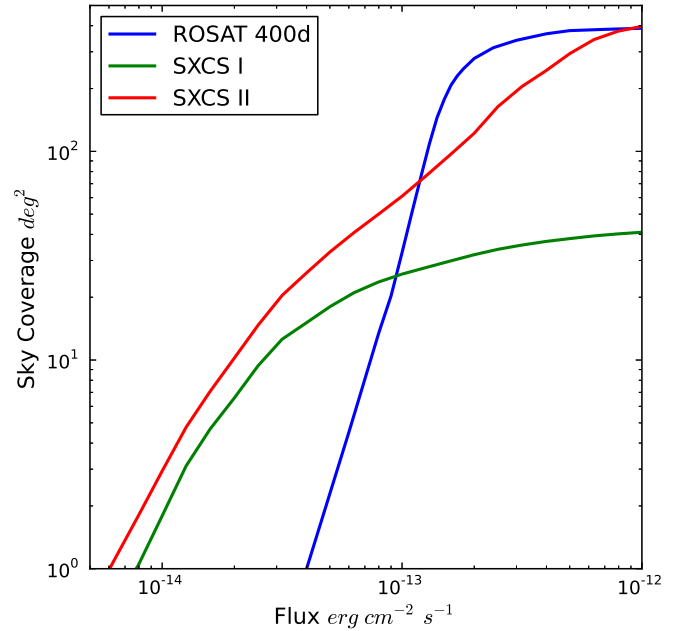}
\caption{Sky coverage of the complete SXCS (Liu et al. in preparation)
  compared with that of the first SXCS release \citep[][]{tundo12} and
  with the 400sd survey \citep{burenin07_400d}.}
\label{skycov}
\end{figure}

The completeness  curve can be used to set a net photon threshold to the final source list.  For example, if one allows a minimum completeness level of 90\%, all sources with more than 80 net photons are included in the final catalog.  The sharp limit in net photons immediately translates into a flux limit for each exposure time.  
For each field we computed a flux-limit map obtained as ECF$(N_H)\times80/Expmap(t)$,  where ECF depends on the Galactic column density, and $Expmap(t)$ (the exposure map in units of effective time) includes the effect of vignetting.
Then, the solid angle covered by the survey above a given flux was obtained by measuring the total solid angle where the flux-limit is lower than a given flux.  
In Figure \ref{skycov} we show the resulting sky coverage $\Omega (S)$ of the total SXCS (Liu et al. in preparation) and compare it with the sky coverage of the 400 Square Degree ROSAT PSPC Galaxy Cluster Survey by \citet{burenin07_400d} and with the sky coverage used in \citet[][]{tundo12}, which corresponds to a sharp limit of 100 net photons and a much smaller number of fields.

\begin{figure}[htbp]
\includegraphics[width=\linewidth]{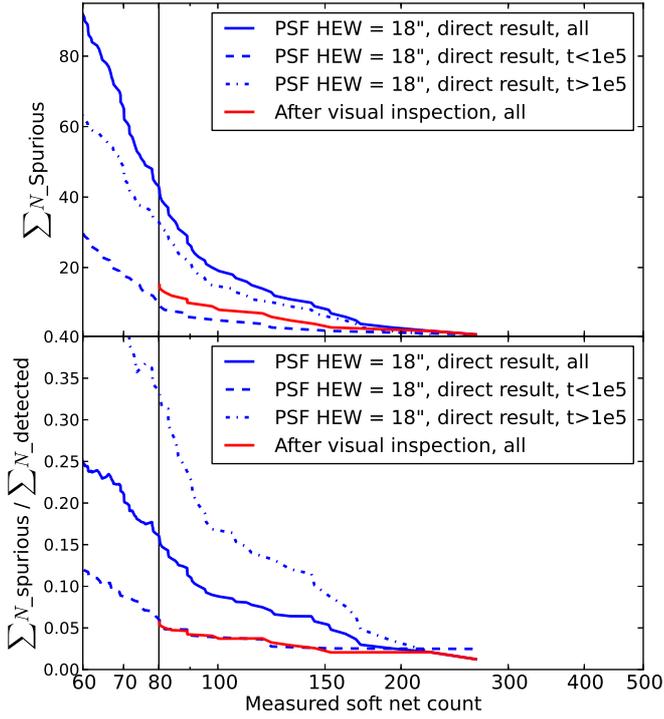}
\caption{Cumulative number (top) and fraction (bottom) of
    sources spuriously classified as extended as a function of the
    measured net photons, before (blue) and after (red) visual
    inspection.  Results of fields with exposure time $< 10^5$ s and
    $> 10^5$ s are plotted separately, showing that most of the
    contamination comes from deep fields.  }
\label{contamination}
\end{figure}

The other key aspect to be kept under control is the contamination level.  The cumulative number of unresolved sources spuriously classified as extended above a given photometry is shown in Figure \ref{contamination}.  This value is rapidly increasing below 100 net counts.  At around 80 net photons, the contamination is reaching a high level of about 40 sources in the entire survey.  As shown in Figure \ref{contamination}, most of the contamination is due to deep fields (with exposure times $> 10^5$ s), which have a higher background and more blended sources.  One option to reduce the contamination is to apply tighter constraints on the source classification.  Of course, this would affect the completeness, and the best compromise must be fine-tuned with several trials.  Another option is to proceed with a visual inspection of all extended source candidates.

Visual inspection has already been used in \citet[][]{tundo12}, and it is often used in {\sl XMM-Newton} cluster surveys as well.  It consists of a careful image inspection of each extended source candidate to reject all sources that appear to be dominated by the contribution of unresolved sources.   Since most of the spurious candidates are found in high-exposure fields, this procedure is performed only on fields with exposure times $> 10^5$ s.  As shown in Figure \ref{contamination} by comparing the solid and dashed lines, this procedure reduces the number of spurious sources with more than $\sim 80$ photons  to an acceptable level: only $\sim$ ten spurious sources are now expected in the entire survey (about $400$ deg$^2$), as opposed to $40$ before the visual inspection.  
In our simulations,  only one  extended source with more than 80 net photons has been erroneously discarded with the visual inspection.
This result shows that X-ray astronomy would greatly benefit by visual analysis, as already successfully done for optical extragalactic astronomy in the form of ``crowd sourcing'' experiments \citep[see, e.g., ][]{lintott08}.

\begin{figure}[htbp]
\includegraphics[width=\linewidth]{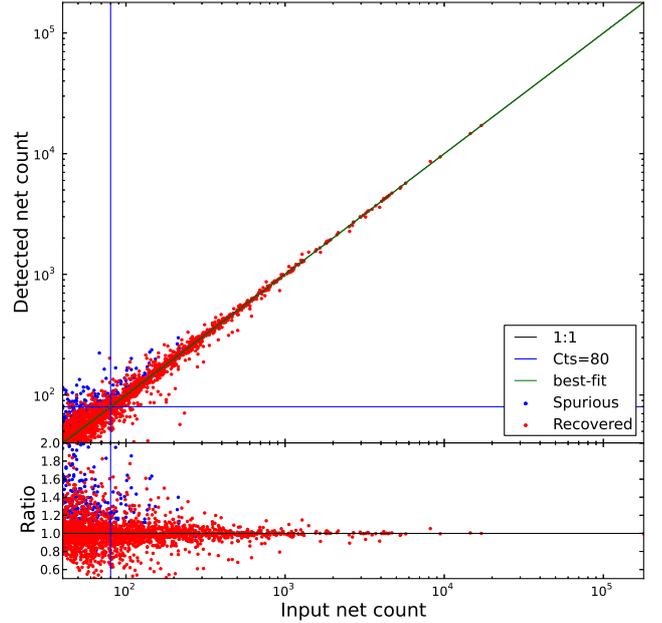}
\caption{Net recovered photons vs input photons for the extended
  sources detected in the simulation (red points).  Spurious extended
  sources are also shown (blue points).  The best-fit for sources
  with more than 80 net photons is $C_{out} = 0.996\times C_{in} +
  0.383$.  }
\label{cts-cts}
\end{figure}

We also  verified the accuracy of our extended source photometry.  
In Figure \ref{cts-cts} we show the input versus the recovered net photons in the soft band for all sources identified as extended.  Spurious extended sources (marked as blue points) typically have overestimated fluxes, because they are often due to blending of unresolved sources.  For all properly characterized extended sources the photometry agrees very well with the actual fluxes and no significant bias is observed.  
The cumulative number counts for sources detected with more than 80 net photons is finally computed as

\begin{equation}
N(>S) = \Sigma_{S_i>S} C_i^{-1}/\Omega(S_i) \, ,
\label{lnls}
\end{equation}

\noindent
where $S$ is the total soft flux, $S_i$ is the soft flux within the extraction radius of the $i^{th}$ source, and $\Omega(S_i)$ is the sky coverage corresponding to $S_i$ \citep[see][]{tundo12}.  
Finally, $C_i$ is the average completeness factor corresponding to a given source photometry, which depends on the net detected photons of the $i^{th}$ source and it is estimated on the basis of the mix of surface brightness distribution assumed in the calibrating simulations (see Figure \ref{completeness}).  In this way, each source is weighted with a factor inversely proportional to the survey completeness.

This treatment is correct only if we can ignore the effects of the statistical errors on the flux measurements.  In reality, flux errors, due to the Poissonian noise on the aperture photometry of each source, introduce a Malmquist bias. In other words, the measured cumulative number counts $N(>S)$ are biased high by an amount roughly estimated as $\Delta N(>S) \sim \alpha \times N(>S) \Delta Cts/ Cts$, where $\alpha$ is the slope of the logN-logS, and $Cts$ is the typical number of measured photons at the survey limit.  
If we assume a low background, as in the case of XRT, we can assume $\Delta Cts \sim \sqrt{Cts}$.  
For $Cts\sim 80$ and $N\sim 300$ as in our case, we estimate $\Delta N\sim 30$, which implies about 35 sources above the detection limit due to the Malmquist bias.  This is a crude estimate, but the Malmquist bias is consistently included in our simulations, since it can be directly computed from the ratio of the recovered and input sources as a function of the {\sl  measured} total net photons.  Once the Malmquist bias was taken into account, we checked that we recovered the input model
of the logN-logS with good accuracy.

\section{Improving the angular resolution}

\begin{figure}[htbp]
\includegraphics[width=\linewidth]{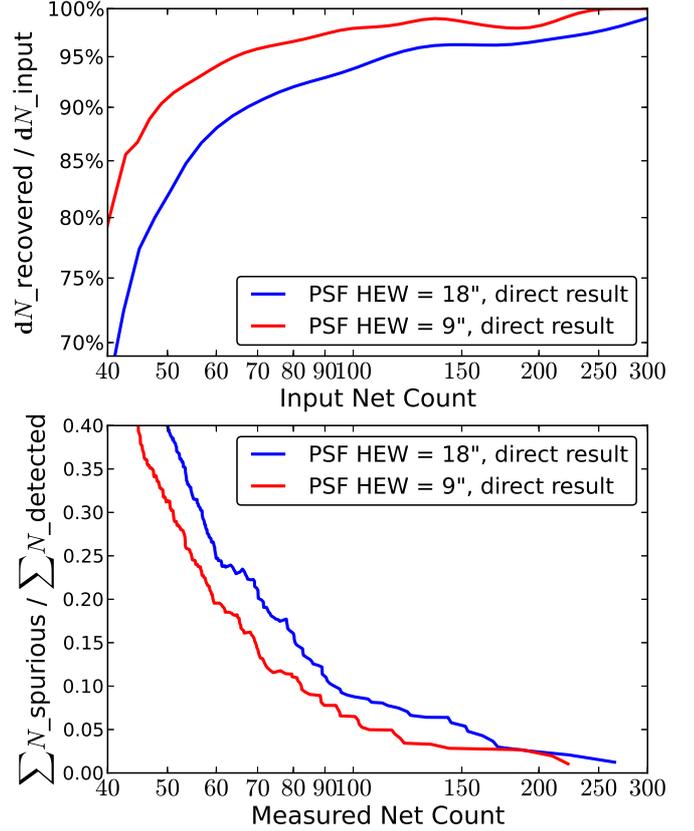}
\caption{Distributions of differential recovered fraction
    (upper panel) and cumulative contamination fraction (lower panel)
    in our simulations, obtained with standard XRT PSF ($HEW = 18$
    arcsec, blue line) and with a sharper PSF ($HEW = 9$ arcsec, red
    line).   }
\label{sharp}
\end{figure}

 Angular resolution is a key  factor for detecting faint sources.
A high angular resolution enables the identification of unresolved sources even with an extremely low number of photons \citep[about 5 for an angular resolution of $1$" -- $90\%$ encircled energy width, so far achieved only by {\sl Chandra} at the aimpoint, see ][]{broos}.  
Angular resolution is clearly also crucial to distinguish between extended and unresolved X-ray sources, and to remove the contribution of faint unresolved sources, which otherwise would be confused with the extended emission itself.
To study the impact of angular resolution on X-ray surveys and in
particular in the identification of extended sources, we again ran the
entire set of simulations after improving the PSF by a factor of two.  This
corresponds to an
imaginary instrument with the same properties as {\sl Swift}-XRT, including
pixel size,  field of view, and  background level, but its PSF
 HEW is only $9$ arcsec, half of the original HEW of XRT \citep{2007Moretti}.  
With this sharper PSF, the core radius was chosen to be 3 pixel sizes ($\sim7\ $ arcsec, enclosing $\sim 65\%$ of the
total source energy of an unresolved source).  
A classification threshold curve was chosen with the same criterion.
As shown in Figure \ref{sharp}, the completeness of the simulation is
significantly improved, while the contamination is reduced.  With this
enhanced resolution, we can identify extended sources down to a much
lower flux limit at similar completeness and contamination levels.
We did not include the visual inspection here, but instead
compared the direct outcome of the source detection algorithm.  As a rule of thumb, we find that by improving the angular resolution by a
factor of two we can identify extended sources down to a flux limit
about 60\% lower for the same completeness and contamination levels.
We also expect that visual inspection would  improve the final results significantly.

This exercise shows that angular resolution is critical for the performance of an X-ray survey mission.  
While a powerful, small FOV, pointing telescope should maximize angular resolution at the aimpoint as in the classical Wolter type mirrors, a survey telescope should maximize the area-weighted angular resolution on the entire FOV.  
The only mission designed according to this criterion is the proposed Wide Field X-ray Telescope \citep[WFXT, ][]{2010WFXT,wfxt1}.  The advantage with respect to a Wolter type-I mirror configuration is clearly shown by the comparison of the $1$ deg$^2$ {\sl Chandra} COSMOS image, worth of the 1.8Ms \citep{elvis2009}, to the simulated WFXT COSMOS field, obtained with only 13ks WFXT exposure \citep{wfxt2}.  The simulated WFXT image has an angular resolution lower by only a factor of $2$ than the average resolution of the {\sl Chandra} mosaic, in front of a $\sim 10$ time  larger HEW of the WFXT PSF with respect to {\sl Chandra}.  
This is because using a Wolter type-I telescope in survey mode results in an average resolution much poorer than the nominal aimpoint resolution, due to the rapid degradation of the angular resolution at large off-axis angle, where most of the FOV resides. 

\section{Conclusions\label{conclusions}}

The combined use of VT and FOF can provide  an efficient algorithm for detecting faint extended sources in X-ray images.   
Although such an algorithm is available to the scientific community for many years \citep[][]{ebeling93}, it has not been used widely for searching extended sources, possibly because of the complex implementation of the method. 
In this work we present an updated implementation of VT and FOF, plus an automated deblending procedure, in a software ({\sl EXSdetect}) for the identification of extended source detection in X-ray images. The aim is to provide a user-friendly, end-to-end algorithm that can be used to exploit present X-ray data archives and to explore the performance of future X-ray missions.  Among the most relevant properties of {\sl EXSdetect} are the following:

\begin{itemize}

\item we include source detection, classification and photometry,
  which are usually performed independently, in a single
  stand-alone algorithm.  We also include a self-consistent
  deblending procedure that efficiently reduces the number of blended
  sources;

\item thanks to our python implementation of the sweep-line algorithm,
  the adopted Voronoi construction scheme is fast also on large
  images;

\item only the PSF of the instrument across the FOV is necessary to run the algorithm,  and no {\sl a
    priori} assumption on the shape of the extended sources is needed;

\item the reasonably short computation time allows one to run
  extensive simulations to optimize the internal parameters
  such as the selection curve for detecting and characterizing
  extended sources, which depend on the survey properties.

\end{itemize}

We also tested our algorithm on extensive simulations run for the SXCS survey.  
While the survey presented in \citet[][]{tundo12} consists of about 40 deg$^2$, here we considered a substantial extension of the survey, with a huge increase of the sky coverage particularly at bright fluxes.  
We find that, with our algorithm, we can detect and characterize extended sources with total net photons as low as $80$ (in the soft band) with a completeness higher than  $90\%$.  
The contamination amounts to a few tens of sources, but can be drastically reduced by visual inspection.  
In the simulations we recovered the input logN-logS with great accuracy down to a flux of $10^{-14}$ erg cm$^{-2}$ s$^{-1}$.  This algorithm is currently being applied to the real SXCS data and the final catalog will be presented in Liu et al. (in preparation).  
The simple exercise of improving the angular resolution by a factor of two shows that the corresponding sensitivity to extended source detection improves by a factor $\sim 1.6$.

To summarize, {\sl EXSdetect} is a new tool to exploit the huge X-ray archives from existing X-ray facilities like {\sl Chandra}, {\sl XMM-Newton}, {\sl Swift}, and {\sl Suzaku}.  At the same time it is also very useful to explore the capability of future X-ray facilities.  The Python code {\sl EXSdetect} is available on the SXCS website (\url{http://adlibitum.oats.inaf.it/sxcs}) and it is open to continuous refinements and updates.

\acknowledgements

We thank the referee, Harald Ebeling, for a detailed and helpful report that significantly improved our work.
LT \& WJX acknowledge support from Chinese NSF grant 10825312 \& 11233002.
We acknowledge support from ASI-INAF I/088/06/0, ASI-INAF I/009/10/0, and from INFN PD51.  

\bibliography{voronoi}

\appendix

\end{document}